\documentclass[12pt]{article}
\usepackage{epsfig}
\def    \mn     {\mu \nu}
\def    \lm     {\lambda}
\def    \lf     {\left (}
\def    \rt     {\right )}
\def    \d      {\delta}
\def    \pl     {\partial}
\def    \t      {\tilde}

\def    \D      {\hat{\Delta}}
\def    \half   {{1 \over 2}}
\def    \comma  {\; , \; \;}
\def    \beq    {\begin{equation}}
\def    \eeq    {\end{equation}}

\begin{document}

{\hfill    SPIN-2000/34 }

{\hfill   {\tt hep-th/0012231}} \vspace*{2cm} \\
\begin{center}
{\Large\bf de Sitter Brane Gravity: from Close-Up to Panorama}
\end{center}
\vspace*{0.5cm}
\begin{center}
{\large\bf Maulik K. Parikh,${}^{\star}${}\footnote{e-mail: 
{\tt m.parikh@phys.uu.nl}}
and Sergey N. Solodukhin${}^{\dagger}${}\footnote{
e-mail: {\tt soloduk@theorie.physik.uni-muenchen.de}}}\\ 
\vskip 1truecm
${}^{\star}$ {\it Spinoza Institute, Utrecht University\\
Leuvenlaan 4, 3584 CE Utrecht, The Netherlands}
\vskip 1.5truemm
${}^{\dagger}$ {\it Theoretische Physik,
Ludwig-Maximilians Universit\"{a}t, \\
Theresienstrasse 37,
D-80333, M\"{u}nchen, Germany}
\end{center}
\vspace*{1cm}
\begin{abstract}
We find explicitly the induced graviton propagator on de Sitter branes
embedded in various five-dimensional spacetimes; de Sitter branes in
AdS and Minkowski space are particular cases. By studying the structure of
the momentum-space propagator, we are able to extract interesting physics,
much of which is qualitatively different from that of flat branes.
We find that 1) there can be a set of graviton-like particles which
mediate brane gravity at different scales; 2) localized gravity can exist
even on de Sitter branes in Minkowski space; 3) Kaluza-Klein modes also
contribute to conventional 4-D gravity for de Sitter branes in AdS; and
4) Newton's constant can vary considerably with scale. We comment on the
implications for the effective cosmological constant.
\end{abstract}

\newpage

\section{Introduction}

The idea that our familiar four-dimensional world might be a slice of a
higher-dimensional spacetime in which the extra dimensions are large
\cite{nima,TeV}, or even infinite \cite{RS}, has generated tremendous interest
and activity. While gauge fields can be trapped
by branes or domain walls, the key finding
of Randall and Sundrum (RS) \cite{RS} 
was that gravity, too, could be localized. Later,
a very intriguing model was proposed by Gregory, Rubakov, and
Sibiryakov (GRS) \cite{GRS} in which gravity is localized only on
intermediate scales and is again five-dimensional at ultra-large scales.
Intermediate scale gravity on the brane is mediated not by
a normalizable graviton state, as happens in the RS model, 
but by a resonant state \cite{CEH,DGP}.
Several variants of such ``quasi-localized gravity'' scenarios are now
known \cite{kogan}-\cite{KR1}.

Most brane-world studies have dealt with Ricci-flat branes. One might
expect new phenomena in such theories when this restriction 
is dropped; indeed, it has even been suggested  \cite{DGP,W,zurab} that the 
cosmological constant problem could be understood within a quasi-localized 
gravity setting. Here we consider de Sitter branes. (Earlier articles 
on de Sitter branes include \cite{Garriga}-\cite{KR}.)
For analyzing the effective brane gravity, it is often more
convenient, in curved space, 
to study the graviton propagator rather than the Newton potential.
By examining the pole and resonance structure of the propagator in its 
momentum-space representation, we are able to understand some of the
scaling behavior of gravity in (quasi-)localized theories. 

In this paper, we shall consider the de Sitter brane
extensions of the RS and GRS scenarios, as follows. First we set
up the geometry. Then we review the technique for 
calculating the momentum-space effective propagator on the brane, with
a brief discussion on how interesting physics can be read off the
propagator. Then, after considering the flat-brane GRS model, we 
determine the de Sitter brane propagator. The expectation of new physics
is indeed borne out. Among the results, we find that Kaluza-Klein modes
also contribute to standard 4-D gravity on de Sitter branes in AdS
(Eq. (\ref{propagator1})), and that gravity is localized even on a de
Sitter brane in flat space (Eq. (\ref{flat})). Most intriguingly, we find
multiple poles and resonances (Eqs. (\ref{res1},\ref{res2})) 
for more general set-ups.
Since each of these graviton-like excitations dominates at a different
energy, we have a picture of gravity at different scales behaving in
drastically different ways, with different Newton couplings that are 
sensitive to the parameters of the configuration. We end with some comments 
on the effective cosmological constant.

Throughout, we focus on the transverse-traceless modes of
the graviton, leaving aside the tensor structure, the radion, questions of 
stability and phenomenological viability, 
and other considerations. Indeed, the GRS model has some drawbacks pertaining
to some of these issues; we work with it here primarily as a tractable toy
model for quasi-localized gravity.

\section{The Set-Up}
 
Consider a visible brane at $y = y_0$ with a hidden brane at $y = y_1 > y_0$, 
with AdS in between, and Minkowski space for $y > y_1$. 
As usual, the whole configuration is 
$Z_2$-symmetric about the visible brane and we consider 
only the half of the spacetime to the right of the visible brane.
The metric takes the form
\beq
ds^2 =dy^2 + e^{-2A(y)}\t{g}_{\mn}(x)dx^{\mu}dx^{\nu} \; .
\eeq
Both AdS and Minkowski space admit constant-curvature foliations,
\beq
\t{R}_{\mn} = 3 \kappa \lm ^2 \t{g}_{\mn}       \; , 
\eeq
where $\kappa=1$ for de Sitter foliations and 
$\kappa=0$ for flat space foliations, with $\lm$ an arbitrary
dimensionful constant.
All quantities with a tilde are derived from $\t{g}_{\mn}$.
The five-dimensional Ricci tensor is zero for Minkowski space and
proportional to the metric for AdS, with a proportionality constant of
$-4k^2$. We choose $\lm = k >0$ without loss of generality.
It is convenient to introduce a new coordinate $z$:
\beq
{dz\over dy} = k e^{A(y)} \; .
\label{z}
\eeq
For a brane placed at a point
$y~(z)$, the Israel junction condition relates the tension to the jump in
the derivative of the warp factor:
\beq
\tau=[\pl_y A(y)] = k e^{A(z)} [\pl_z A(z)] \; . \label{tension}
\eeq
We shall consider perturbations along the brane. 
The perturbed bulk metric reads
\beq
ds^2=dy^2+\lf e^{-2A(y)}\tilde{g}_{\mn}(x)+h_{\mn}(x,y) \rt dx^{\mu}dx^{\nu}
\; , \label{metric1}
\eeq
where the graviton $h_{\mn}(x,y)$ is taken to be traceless and to 
satisfy the gauge choice $\nabla_\mu h^{\mn} = 0$.

\section{The Propagator}

In this paper we are interested in the graviton propagator for the 
perturbations in Eq. (\ref{metric1}).
The five-dimensional propagator for the transverse-traceless graviton 
is the solution to the equation
\beq
\lf \nabla_5^2 + 2k^2 \rt \Delta^{\alpha' \beta'}_{\mn} (x, y; x', y') 
= G_5 \d (y - y') {\d^4 (x - x') \over \sqrt{-g(x)}} 
\d ^{\alpha'}_{\mu} \d^{\beta'}_{\nu} \; ,
\eeq
where $G_5$ is the five-dimensional Newton constant. The coordinate-space
propagator can be computed directly via an integral over the 
appropriately-normalized Kaluza-Klein and zero modes of the brane. However,
this results in a formidable integral of a hypergeometric function. Instead, 
we expand the graviton propagator as an integral
in the Fourier modes of the brane:
\beq
\Delta^{\alpha' \beta'}_{\mn} (x, y; x', y') = \int {d^4 \t{p} \over 
\lf 2 \pi \rt ^4} f^{\alpha' \beta'}_{\mn} (\t{p}, x, x') 
\Delta_p (y, y') ~e^{4A(y_0)} \; , \label{fourier}
\eeq
where the Fourier modes are defined to satisfy
\beq
\nabla_{\t{g}}^2 f^{\alpha' \beta'}_{\mn} (\t{p}, x , x') 
= - \t{p}^2 f^{\alpha' \beta'}_{\mn} (\t{p}, x , x') \; .
\eeq
The factor $\exp(4A(y_0))$ in Eq. (\ref{fourier}) has been included because
Eq. (\ref{fourier}) is an
expansion over the $\t{p}$-momentum modes. We are ultimately interested, 
however, in the physical momentum modes, $p$, of the visible brane.
The two sets of modes are related via $p^2 = \t{p}^2 \exp^{+2A(z_0)}$. By
accounting for this factor, we ensure that $\Delta_p (y,y')$ is in terms
of physical momentum.

The analysis for the effective graviton propagator on a brane embedded in
a warped compactification scenario was carried out by Giddings, Katz, and
Randall (GKR) \cite{GKR}. Here we review the aspects of that work that
will be useful later. Defining the momentum-space propagator as
\beq
\Delta_p (y, y') \equiv e^{-A(y)/2} \D_p (z, z') e^{-A(y')/2} ~e^{-4A(y_0)}\; ,
\label{DD}
\eeq
we find that
\beq
\lf \pl_z^2 - V(z) - \lf {\t{p} \over k} \rt^2 \rt \D_p(z,z') 
= {G_5 \over k} e^{+4A(z)} \d (z - z') \; ,     
\label{Schred}
\eeq
where the potential $V(z)$ is
\beq
V(z) = {17 \over 4} \lf \pl_z A \rt^2 + {1 \over 2} \pl_z^2 A \; .
\eeq
For $z \neq z'$, $\D_p(z,z')$ satisfies an analog 
Schr\"{o}dinger equation, Eq. (\ref{Schred}). 
We now define two functions: $\D_{<}(z,z')$ if $z\leq z'$
and $\D_{>}(z,z')$ if $z\geq z'$. The general solution
thus has four constants of integration in each region of spacetime.

These constants are fixed by an equal number of conditions.
At $z = z'$, Eq. (\ref{Schred}) and a matching condition imply that
\beq
\D_<|_{z=z'} = \D_>|_{z=z'} \comma \pl_z(\D_>-\D_<)|_{z=z'} 
= {G_5 \over k} e^{+4A(z)} \; . \label{bcon}
\eeq
Additionally, the junction conditions at each brane give
\beq
[\D]|_{\rm brane} = 0 
\comma [\pl_z \D]|_{\rm brane} = -{3 \over 2}[\pl_z A(z)] 
\D|_{\rm brane} \; . \label{jump}
\eeq
Finally, we have to specify the boundary condition at infinity. We require
that there be only outgoing modes there; nothing comes in from infinity.

In the remainder of this paper, we will solve for the effective propagator
on the visible brane by setting $z = z'= z_0$. From Eq. (\ref{DD}), the 
propagator as a function of physical momentum on the brane, is
\beq
\Delta (p) = \D_p \lf z_0, z_0 \rt e^{-5A(z_0)} \; .    \label{scaling}
\eeq
Much can be understood from the form of the momentum-space graviton
propagator. The propagator for the transverse-traceless modes of a graviton
on de Sitter space of radius $R_0$, when
written in terms of $q^2 \equiv - p^2$, takes the form
\beq
\Delta(q) = {G_N \over q^2 - 2 R_0^{-2}} \; . \label{desitter}
\eeq
Hence, for our effective brane propagator, a pole in $q^2$ indicates the 
presence of a four-dimensional graviton on the visible brane; 
the residue gives the effective four-dimensional Newton constant.
A pole at negative $q^2$ indicates that Kaluza-Klein modes conspire to give a
graviton-like resonance. 
On the other hand, an effective propagator that goes as $1 / q$
indicates that gravity on the brane 
is still five-dimensional; Gauss' law is not obeyed on the brane. 
Also, since there are always KK modes, the propagator
has in general an imaginary part which is related to the
flux of gravitational radiation into the bulk. And finally,
the effective cosmological constant can be read off from the 
inverse of the momentum-space propagator in the limit of zero momentum.

\section{Flat Branes}

Although we are mainly interested in de Sitter branes,
we pause here to consider the original flat-brane 
GRS model. The function $A(y)$ is $ky$ for AdS region and is a constant,
$ky_1$, in the Minkowski region. The visible brane is at $y=0$ and has tension
$\tau_0=2k$ while the hidden brane  is at $y=y_1$ and its tension $\tau_1=-k$
is negative. The basis of solutions of the analog Schr\"{o}dinger equation,
Eq. (\ref{Schred}), for this case was analysed in GRS \cite{GRS}
and consists of Bessel functions. One finds that the propagator along 
the brane $(y=y'=0)$, when written in terms of $x = q/k$, is
\beq
\Delta (x)={G_5  \over kx} \lf \alpha J_2(x)-\beta N_2(x)\over
\alpha J_1(x)-\beta N_1(x) \rt \; ,
\label{D}
\eeq
where
\begin{eqnarray}
&&\alpha=i N_2(xz_1)-N_1(xz_1) \nonumber \\
&&\beta=i J_2(xz_1)-J_1(xz_1) \; ,
\label{ab}
\end{eqnarray}
with $z_1 = e^{ky_1}$.
It is instructive to analyze Eq. (\ref{D}) for small values of $x$. 
Expanding the numerator and
denominator in powers of $x$, we get
\beq
{\rm Re} ~ \Delta (x)={2 G_5  \over k z_1^2}{z_1^2+{1\over 2} \over
x^2+ \lf {2\over z_1(z_1^2-1)} \rt ^2}
\label{Re}
\eeq
\beq
{\rm Im} ~ \Delta (x)=-{4 G_5 \over k} {z_1 \over \lf z_1^2-1 \rt ^2}{1\over x}
{1 \over x^2+ \lf {2\over z_1(z_1^2-1)} \rt^2}  \; .
\label{Im}
\eeq
The imaginary part of $\Delta (q)$ has a pole at $q=0$ near which the
propagator is
\beq
\Delta (q)\simeq {G_5 \over i q} z_1^3 \; .
\label{propq}
\eeq
This is a reflection of the fact that gravity on a flat 
brane becomes effectively five-dimensional at large distances.
On the other hand, the structure of ${\rm Re}~ \Delta (q)$ establishes the 
presence of a resonance state in the spectrum. The real part of the propagator 
is thus effectively four-dimensional with an effective Newton constant of
\beq
G_N= 2 G_5 k \lf 1 + {1 \over 2 z_1^2} \rt 
\stackrel{z_1 \to \infty}{\simeq} 2 G_5 k \; .
\eeq
Thus, for flat branes, the effective brane gravity is four-dimensional
at intermediate scales but five-dimensional at very small or very large scales.
As we shall see, on de Sitter branes things are rather different.

\section{de Sitter Branes}

Now consider de Sitter branes. 
The warp factor for de Sitter-foliated AdS and flat space is
\begin{eqnarray}
&& \left . e^{-2A} \right |_{\rm AdS}
= \sinh^2(-ky)={1\over \sinh^2z} \nonumber \\
&& \left . e^{-2A} \right |_{\rm FS} = k^2 (y-b)^2=e^{2(z+c)} \; ,
\label{warp}
\end{eqnarray}
where $b$ ($c$) is a constant determined by joining the two spaces 
at $y=y_1$ ($z=z_1$). For AdS, $y=0$ corresponds to the AdS Cauchy horizon; 
regions on different sides of the horizon should be
considered separately\footnote{We contrast this with the set-up of 
\cite{TW} in which configurations with branes on different sides 
of the horizon are considered.}.
We see from Eq. (\ref{warp}) that in AdS and flat spaces
the warp factor has both decaying and increasing branches. There are therefore
four different ways to configure the two spaces, namely: A) AdS and FS both 
have decreasing warp factors;
B) the warp factor increases in AdS but decreases in FS; C) the warp factor 
decreases in AdS and
increases in FS; D) the warp factor increases everywhere. 
Here we will consider mostly configuration A. Specifically,
\beq
e^{A(z)}=
 \left\{
\begin{array}{lc}
\sinh z & z_0\leq z\leq z_1 \\
\sinh z_1 e^{(z-z_1)} & z\geq z_1 \; .
\end{array}
\right. \label{warp1}
\eeq
The tension of the visible and the hidden brane is, respectively,
\beq
\tau_0 = 2k \cosh z_0 \comma \tau_1 = - k (\cosh z_1-\sinh z_1) \; .
\label{tensions}
\eeq
The potential $V(z)$ corresponding to Eq. (\ref{warp1}) is
\beq
V(z)= \left\{
\begin{array}{lc}
{17\over 4} + {15\over 4}{1\over \sinh^2z} & z_0 \leq z \leq z_1 \\
{17\over 4} & z \geq z_1 \; .
\end{array}
\right.
\eeq
By switching to a new variable,
\beq
\xi \equiv + \coth z \; ,
\eeq
the solution to the analog Schr\"{o}dinger equation, Eq. (\ref{Schred}), 
in the AdS region is readily obtained in terms of 
associated Legendre functions, $P^{iM}_{3/2} (\xi)$ and $Q^{iM}_{3/2} (\xi)$.
(Some useful formulas for manipulating these functions are listed in the
appendix.) The parameter $M$ that appears in the superscript is
\beq
M^2 = q^2 R_0^2 - {17 \over 4} \; ,
\eeq
where $q$ is the physical momentum on the visible brane, 
$q = \t{q} \exp (A(z_0))$, $\tilde{q}^2=-\tilde{p}^2$, and $R_0$ is the physical radius of the brane:
\beq
R_0^{-1} = k \exp (A(z_0)) \; .
\eeq

In the flat space region,
the solution is given by a combination of ingoing and outgoing plane
waves, $e^{\pm iMz}$. Together with the 
associated Legendre functions in the AdS region, these form a continuous 
spectrum with $M^2\geq 0$. Analyzing the spectrum with $M^2<0$ one finds that
the only such mode which is both normalizable 
and satisfies all boundary conditions has the form $e^{-{3\over 2}z}$ 
in the flat space region and $P^{-3/2}_{3/2}(\coth z )$ in Anti-de Sitter region
that corresponds to $iM=-{3\over 2}$.
There is thus a gap between the bound state
and the continuous portion of the spectrum \cite{Garriga}. 
On the visible brane the 
bound state corresponds to the four-dimensional graviton,
which has $q^2 = 2R_0^{-2}$.

Next, we define two functions, $\D_<(z,z')$ and $\D_>(z,z')$, 
valid respectively when $z \leq z'$ 
($\xi \geq \xi'$) and $z \geq z'$ ($\xi \leq \xi'$).
Since we will ultimately take $z$ and $z'$ both to be on the visible brane,
we simplify the algebra by choosing $z'$ to be in the AdS region from 
the outset; $z$ may be located in either region.   
After imposing the junction condition at the visible brane 
(using Eq. (\ref{derivative}) from the appendix), the propagator takes the form
\beq
\D^M_< (z, z') = C_M (\xi') \lf Q^{iM}_{1/2} (\xi_0) 
P^{i M}_{3/2} (\xi) - P^{iM}_{1/2} (\xi_0) Q^{iM}_{3/2} (\xi) \rt \; ,
\eeq
and
\beq
\D^M_> (z, z') = \left\{
\begin{array}{lc}
A_M (\xi') P^{i M}_{3/2} (\xi) + B_M(\xi') Q^{iM}_{3/2} (\xi) & z \leq z_1 \\
D_M(z') e^{i M (z-z_1)} & z \geq z_1
\end{array}
\right.
\eeq
where, in accordance with our boundary condition at infinity, only
outgoing modes have been retained in the flat space region.

The other conditions, Eq. (\ref{bcon}) and Eq. (\ref{jump}) at the 
hidden brane,
fix the remaining coefficients $A(\xi')$, $B(\xi')$, $C(\xi' )$, and $D(z')$. 
The expression for the propagator for arbitrary $\xi$ and $\xi'$ is 
quite lengthy.
However, when both points are on the brane ($\xi=\xi'=\xi_0$), the
expression takes a nice form. Using Eq. (\ref{derivative}) and
not forgetting Eq. (\ref{scaling}), a little bit of work gives
\beq 
\Delta (q)= G_5 R_0 {1\over {3 \over 2} + iM}
\lf \alpha (M)Q^{i M}_{3/2}(\xi_0)- \beta (M) P^{i M}_{3/2}(\xi_0) \over 
\alpha (M) Q^{i M}_{1/2}(\xi_0)- \beta (M)P^{i M}_{1/2}(\xi_0)) \rt \; ,
\label{Dprop}
\eeq
where all the $q$-dependence is contained in $M$. Here
\begin{eqnarray}
&&\alpha (M)\equiv P^{i M}_{3/2}(\xi_1)-P^{i M}_{1/2}(\xi_1) \nonumber \\
&&\beta (M)\equiv Q^{i M}_{3/2}(\xi_1)-Q^{i M}_{1/2}(\xi_1) \; .
\label{ab1}
\end{eqnarray}
Eq. (\ref{Dprop}) is the desired formula; it required solving some linear 
algebraic equations for the coefficients, but no integration. As we shall
see, the detailed structure of our propagator contains a wealth of
information.

There are some limiting regimes which are of particular interest.

\subsection{Curved Randall-Sundrum Limit} 

Take the limit when $z_1 \to \infty (\xi_1 \to 1)$. 
Then the radius of the second brane shrinks 
to zero, and the Minkowski region disappears. We are thus in the de 
Sitter-brane Randall-Sundrum limit. Now, from properties of the Legendre
functions, we know that when $\xi_1$ goes to 1, $\alpha (M)$ goes to 0.
Eq. (\ref{Dprop}) then reads
\beq
\Delta (q)={G_5R_0\over {3\over 2}+i M} {P^{i M}_{3/2} (\coth z_0)
\over P^{i M}_{1/2} (\coth z_0)} \; .
\label{RS1}
\eeq
In Fig. 1, we plot the imaginary part of this function.
\begin{figure} 
\centerline{\psfig{figure=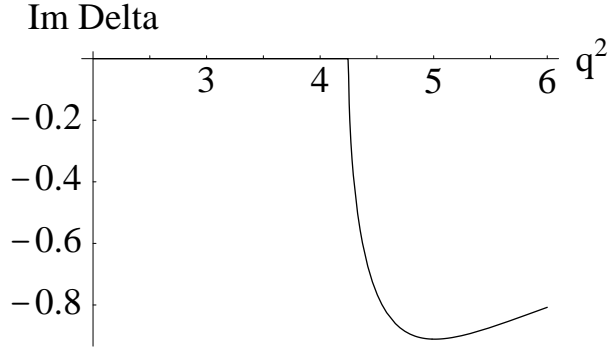}}
\caption{Imaginary part of the propagator at $\xi_0 = 2$, showing the onset of
Kaluza-Klein excitations.}
\end{figure}
We see that the imaginary part of the propagator is non-zero only for 
$M^2 > 0$ and is entirely due to
the continuous (Kaluza-Klein) modes. This is how the gap in the spectrum
shows up in the structure of the propagator. Another manifestation of the gap
is a jump in the derivative of the real part of the propagator at $M=0$.

Moreover, using a Legendre function identity, Eq. (\ref{identity}), we can
rewrite Eq. (\ref{RS1}) as
\beq
\Delta (q) = {2 G_5 k \cosh z_0 \over q^2 - 2 R_0^{-2}} - G_5 k \sinh z_0
 {\half + i \sqrt{{q^2R_0^2} - {17 \over 4}} \over 
q^2 - 2 R_0^{-2}} {P^{i \sqrt{{q^2 R_0^2} - {17 \over 4}}}_{-\half} (\coth z_0) \over
P^{i \sqrt{{q^2 R_0^2} - {17 \over 4}}}_{+\half}(\coth z_0)} \; .
\label{propagator1}
\eeq
The first term above is the contribution of
the zero mode while the second term has a Kaluza-Klein origin.
Eq. (\ref{propagator1}) is similar to the corresponding 
expression obtained in \cite{GKR} for flat branes.
We know from the analysis of \cite{RS} and \cite{GKR} that 
only the zero mode leads to localized 4-D gravity for flat branes;
the Kaluza-Klein modes give a correction. But for de Sitter branes
the story is quite different: both the zero mode and the KK modes 
contribute to the four-dimensional propagator on the brane. 
This can be seen from the fact that both terms in
Eq. (\ref{propagator1}) have a pole at $q^2=2R_0^{-2}$. 
To simplify the analysis we consider the case when $\xi_0 \gg 1$ ($R_0k\gg 1$).
In this regime, using the asymptotic expression Eq. (\ref{ratio}) 
in the appendix, we find that
\beq
\Delta (q)\simeq {2G_5k \over  \lf q^2-2R_0^{-2} \rt }
{\lf 1-{1\over 2k^2}(q^2-4R_0^{-2}) \ln (R_0k)
\rt}  \; . \label{pole}
\eeq
According to our prescription, the induced Newton constant is identified
as the residue of the graviton propagator at the pole $q^2 = 2R_0^{-2}$.
Thus, for $R_0k \gg 1$, we find
\beq
G_N \simeq 2G_5 k \lf 1+(R_0k)^{-2} \ln (R_0k) \rt \; .
\label{GN}
\eeq
The value of the induced cosmological constant (or better to say the 
vacuum energy) 
$\Lambda$ can be read off as
well, using $\Lambda\equiv {G_N\over -2\Delta (q=0)}$. 
Taking the limit $q\to 0$ in Eq. (\ref{pole}) we find
for $R_0 k \gg 1$, that
\beq
\Lambda \simeq{1\over R_0^2}\lf 1 - (R_0k)^{-2}\ln (R_0k) \rt   \; .
\label{Lambda}
\eeq
We mention in passing that since cutoff-AdS/CFT should be applicable to the
curved Randall-Sundrum limit, sub-leading terms in the propagator 
could be interpreted as originating in 
interactions with conformal matter on de Sitter space. 
In particular, the logarithmic term in Eqs. (\ref{GN},\ref{Lambda})
may be related to the conformal anomaly.
 
\subsection{Coincident Branes: de Sitter Brane in Flat Spacetime}

Now consider the opposite limit, in which the two branes are 
coincident: $z_0=z_1$. In this limit the
anti-de Sitter region between the branes disappears, and 
it is as we had a single de Sitter brane in Minkowski space.
This brane has positive tension provided we start with configuration A.
Consulting our master formula, Eq. (\ref{Dprop}), we see that
when $z_0=z_1$ the propagator reduces to a simple form:
\beq
\Delta (q)={G_5R_0\over {3\over 2}+i M} \; .
\label{flat}
\eeq
Evidently, Eq. (\ref{flat}) has a pole at $iM=-{3\over 2}$ where
$q^2 = 2R_0^{-2}$. There is thus four-dimensional gravity on the brane
even in Minkowski space. This is a consequence of the normalizability of the
bound state.
(Had we considered a negative tension brane, for which the factor $e^{-2A}$ 
grows with $z$, the pole (here at $iM={3\over 2}$) would not have
corresponded to a physical particle since there are now no normalizable bound 
states.)
We also obtain the Newton constant:
\beq
\Delta (q) \stackrel{M \to 3i/2}{\simeq} {3G_5\over R_0}\ 
{1\over q^2-2R_0^{-2}} \Rightarrow 
G_N={3G_5\over R_0}      \; .    \label{Newton}
\eeq
This depends on the radius of the brane or, equivalently, on the 
location of the brane in the five-dimensional flat 
spacetime. That localized gravity can arise
on a curved brane in asymptotically flat space with a position-dependent 
Newton constant is a by-product of the recent study in \cite{SKS}.
Note that the effect is absent for flat branes. 
Indeed, the propagator along the brane is then
\beq
\lim_{R_0 \to \infty} \Delta (q)={G_5 \over i q} \; ,
\label{propf}
\eeq
which is a five-dimensional propagator. We note as a check that this is
just Eq. (\ref{propq}) in the limit $z_1 \to 1$.

Turning now to the continuous part of the spectrum, $M^2 \equiv m^2 R_0^2 
\geq 0$, we find that
\beq
{\rm Re}~ \Delta (m)={3\over 2}{G_5\over R_0}{1\over m^2+{9\over 4R_0^2}} 
\comma
{\rm Im}~ \Delta (m)=-{G_5 m\over m^2+{9\over 4R_0^2}}    \; .
\label{resonance}
\eeq
This has the characteristic Breit-Wigner form of a resonance. The resonance is
concentrated at $m=0$ and has a decay width of ${3 \over 2 R_0}$.
In other words, brane gravity is mediated by {\it both} a graviton and a 
graviton-like resonance. Their Newton couplings to matter differ by a factor
of two.

\subsection{Branes at Finite Separation}

A general feature of the graviton propagator for two de Sitter branes
in configuration A is that the flat-brane pole ${1\over iq}$, 
which signals the re-emergence of 5-D gravity at ultra-large scales,
now disappears. It is replaced by a pole at $iM=-{3\over 2}$ 
or, equivalently, at $q^2=2R_0^{-2}$. This
corresponds to the four-dimensional graviton 
which now mediates the large-scale gravitational interaction on 
the brane. The strength of this interaction, Newton's 
constant, is computed by taking
the residue of Eq. (\ref{Dprop}) at the pole $iM=-{3\over 2}$.  
This can be done explicitly by computing the residue\footnote{
This requires some care since the Legendre 
function $Q^{-3/2}_{1/2}$ is not defined: one has to first transform 
$Q^{iM}_{1/2}$ into $Q^{-iM}_{1/2}$ and then take the limit $iM \to -{3/2}$.} 
in Eq. (\ref{Dprop}). The physics simplifies in the
approximation of large radius branes: $R_0k, R_1k \gg 1$. 
In this regime, both $\xi_0$ and $\xi_1$ are large, and the
propagator displays not only the pole but also the rich resonance structure.
Using Eqs. (\ref{asymptP},\ref{asymptQ}) we expand
the Legendre functions in Eq. (\ref{Dprop}) for large $\xi_0$ and $\xi_1$.
We find that\footnote{Note that Eq. (\ref{Daprox})
is not valid for the case of single brane in AdS space since in this case $\xi_1=1$ (the second brane is taken to infinity).}
\beq
\Delta (q) \simeq {G_5R_0\over {3\over 2}+iM} {1 \over \alpha}\lf
{ 4kR_0 \alpha - \lf 1 - \alpha^4 \rt \lf {3\over 2}+iM \rt \over
4kR_0 \alpha^3 +2 \lf 1 - \alpha^2 \rt \lf {3\over 2}-iM \rt}\rt \; ,
\label{Daprox}
\eeq
where we have introduced $\alpha=\lf{R_1\over R_0}\rt$. From
the pole at $iM=-{3\over 2}$, we see that
\beq
G_N^{\rm pole}\simeq {3G_5\over R_0}\lf{R_0\over R_1}\rt^3 \; .
\label{gn}
\eeq
Up to a factor $\alpha^{-3}$, this is similar to the 
expression we have for coincident branes. This is because 
at large intervals the merger of two branes looks like one
de Sitter brane in flat space. The configuration-dependence of 
the four-dimensional Newton constant is evident.

On the other hand, when the brane separation is relatively large we expect a
resonance to appear in the KK portion of the spectrum, similar to the
resonance in flat-brane GRS.
In fact, Eq. (\ref{Daprox}) contains not one, but two resonances!
To see this, let $M=mR_0$ and $\alpha^3 \xi_0 \gg 1$. Then
\beq
\Delta (q) \simeq \Delta_1(q)+\Delta_2 (q) \; ,
\eeq
where
\beq
\Delta_1(q)=G_5{1\over \alpha^3}\ {{3\over 2R_0}-im\over {9\over 4R_0^2}+m^2} \; , \label{res1}
\eeq 
and
\beq
\Delta_2(q)=G_5{(1-\alpha^2)(2+\alpha^2)\over 2\alpha^3}\ {({3\over 2R_0}+2k{\alpha^3\over 1-\alpha^2})+im
\over ({3\over 2R_0}+2k{\alpha^3\over 1-\alpha^2})^2+m^2} \; .
\label{res2}
\eeq 
Both  Eq. (\ref{res1}) and Eq. (\ref{res2}) have the resonance 
structure familiar from the flat GRS case, cf. Eqs. (\ref{Re},\ref{Im}).
The gravitational coupling is, respectively,
\beq
G^{\rm res1}_N={3\over 2}{G_5\over R_0}{1\over \alpha^3} \comma
G_N^{\rm res2}=kG_5(2+\alpha^2) \; .
\label{Gres}
\eeq
We see that the second resonance is analogous to the resonance in flat-brane
GRS. The coupling $G_N^{\rm res2}\sim 2kG_5$ (for small $\alpha$) is typical for a 
single brane embedded in anti-de Sitter space.
The first resonance is what one could call a {\it shadow},
since it always accompanies the pole at $iM=-{3\over 2}$. 
We have already seen it at the end of the previous section, in Eq. 
(\ref{resonance}).
In the flat brane limit when $R_0$ goes to infinity 
(with $\alpha={R_1\over R_0}$ fixed) the second resonance becomes
the GRS resonance while the ``shadow'' disappears. But not completely: 
its imaginary part $- i G_5 \alpha^{-3}/ m$ survives.
The small $x$-divergence of ${\rm Im~} \Delta (x)$,
in Eq. (\ref{Im}) is all that is left of the ``shadow.''
From the real part of the propagator, Eqs. (\ref{res1},\ref{res2}), 
we read off the widths of the resonance:
\beq
\Delta m_1={3\over 2R_0} \comma
\Delta m_2={3\over 2R_0}+2k{\alpha^3\over 1-
\alpha^2}\simeq 2k\lf{R_1\over R_0}\rt^3 \; .
\eeq
We see then that we have a ``zoo'' of graviton-like states -- a pole and
two resonances -- which mediate 4-D gravity on the brane. 
We will discuss the physical consequences of this 
in the next section. Here we just mention that the  gravitons
operate at essentially different scales. The second resonant state
is responsible for the gravitational interaction at intervals
\beq
\Delta s_2\sim {1\over \Delta m_2}\simeq {1\over 2k} \lf{R_0\over R_1}\rt^3
\; ,
\eeq
while the pole becomes important at much larger intervals, 
$\Delta s\gg\Delta s_2$.
The  ``shadow'' operates somewhere 
in between. It seems that its role is just to make 
a smooth  transition  between two completely different regimes.

\section{From Close-Up to Panorama}

In this paper, we have calculated the effective momentum-space 
graviton propagator on a de Sitter brane embedded in 
various five-dimensional spacetimes. The resulting picture is of gravity
with rich scale structure. Not only does the approximate 
dimensionality of Newton's law depend on the scale, but the coupling and
even the identity of the carrier of the gravitational interaction -- whether
the graviton or a resonance -- depends on the
energies at which the physics is probed.

This is because the gravitational interaction on a brane 
propagates not just along the brane but also
through the bulk. The larger the separation between points on the brane,
the deeper the part of the bulk that affects the propagation. 
At very short distances on the brane,
gravity is five-dimensional. This is due to large-momentum
excitations; it can be seen by taking the limit of infinite $q$ in any
one of our propagators. The brane propagator then behaves as ${1\over iq}$;
that is, as a five-dimensional graviton. 
The scale at which the physics is five-dimensional is
of order $1/k$. 

When a configuration of two branes is considered, the brane separation 
sets another scale (see \cite{GRS}) at which the gravitational interaction 
on the brane can change dramatically. This scale, $\Delta s_2\simeq {1\over 2k}({R_0\over R_1})^3$, is determined by the 
width of the resonant mode
which mediates the four-dimensional gravitational force in the 
range $1/k\ll\Delta s \ll\Delta s_2$.
For well-separated branes $\Delta s_2$ 
can be much larger than $1/k$. In order to make estimates let us assume that 
$R_0k , R_1k \gg 1$.
Then the radius of a brane can be approximated as $R={1\over k}e^{ky}$, 
where $y$ is the geodesic distance between the brane and the AdS horizon. 
The strength of the gravitational interaction
for the above range of intervals is given by 
$G^{\rm res2}_N$, Eq. (\ref{Gres}), which is approximately
$2kG_5$. This is because for the intervals $1/k\ll\Delta s\ll\Delta s_2$,
the second brane has no effect.

For intervals larger than $\Delta s_2$ the propagation through
the bulk is affected by the second brane and Minkowski space; for  
$\Delta s\gg\Delta s_2$ propagation is mostly through flat space. 
Gravity on the brane in this regime
is mediated by the pole with Newton constant $G^{\rm pole}_N$, Eq. (\ref{gn}).
The interaction then resembles that of a single 
brane embedded in flat space. Comparing
Eq. (\ref{gn}) and Eq. (\ref{Gres}) we see that gravity
at larger scales is much weaker than at short scales:
\beq
G_N^{\rm pole}/G_N^{\rm res2}\simeq {1\over kR_0}\lf{R_0\over R_1}\rt^3 \; .
\label{2}
\eeq
For $R_0 \sim R_1$ this is exponentially small.
The brane cosmological constant (or vacuum energy) is determined,
to leading order, by its geometric value:
$\Lambda\simeq {1\over R^2_0}$. 
For the dimensionless quantity $\Lambda G_N$ we find
\beq
\Lambda G^{\rm res2}_N\simeq {1\over k^2R_0^2}\sim e^{-2ky_0} \; ,
\label{lambgn}
\eeq
provided we choose $G_5\sim k^{-3}$.
But measuring the cosmological constant in terms of 
$G^{\rm pole}_N$ we obtain
\beq
\Lambda G^{\rm pole}_N\simeq {1\over (kR_0)^3}\lf{R_0\over R_1}\rt^3 \; ,
\label{3}
\eeq
which is smaller than that of Eq. (\ref{lambgn}) by a 
factor $1/(kR_0) \sim e^{-ky_0}$, assuming that $y_0\sim y_1$.

These estimates are even stronger if we consider
configuration B. In this case, the radius of the visible brane is smaller 
than that of the hidden brane, $R_0<R_1$, so that the visible brane has 
negative tension. The total tension of the stack of three branes is 
still positive, though.
So at larger scales, gravity on the visible brane is still due to the 
pole, with the Newton constant given by Eq. (\ref{gn}). 
At shorter scales one could expect some pecularities  
due to the negative tension. Indeed, 
the overall factor $(1-\alpha^2)$ in Eq. (\ref{res2}) flips sign 
when $R_1$ becomes greater than $R_0$. However, closer inspection of 
Eq. (\ref{res2}) shows that for $R_0k\gg1$,
the Newton constant $G_N^{\rm res2}$ is given by Eq. (\ref{Gres})
and is still positive. Assuming that $R_1$ is much larger than $R_0$, we find 
from Eq. (\ref{Gres}) that
\beq
G^{\rm res2}_N\simeq 2kG_5 \lf {R_1\over R_0} \rt \; .
\label{4}
\eeq
So for $R_1 \gg R_0$ the Newton constant of the resonance in configuration B 
is much smaller than it is for configuration A.
Comparing Eq. (\ref{4}) with $G^{\rm pole}_N$ we see that
\beq
G_N^{\rm pole}/G^{\rm res2}_N\simeq {1\over kR_0}\lf{R_0\over R_1}\rt^4
\sim e^{-ky_0}e^{-4k|y_0-y_1|} \; .
\label{5}
\eeq
The large-scale gravitational interaction is much 
weaker than at shorter scales
mostly due to the difference between $R_1$ and $R_0$.  
Consequently, the cosmological constant measured in units of $G_N$ is 
exponentially smaller at larger scales.

If the present picture is correct and we live on the visible brane in a
stack of three branes arranged in configuration A or B, then there is a
relation between the
observable Newton constant, $G_N$, the de Sitter radius of the brane, $R_0$,
and the five-dimensional Newton
constant, $G_5$. 
Inverting Eq. (\ref{gn}) for $G_5$ we can estimate the fundamental
energy scale in the bulk, $M^{(5)}_{\rm pl}=G^{-1/3}_5$.  Inserting 
the measured values of $G_N$ and the Hubble radius, we find
\beq
M^{(5)}_{\rm pl}\sim 10\alpha^{-1} \ GeV \; ,
\label{6}
\eeq
which for $\alpha\simeq {R_1\over R_0}\sim 10^{-2}$ is of the order of the 
electroweak energy scale.

Many interesting questions remain. It would be
worthwhile to consider the implications for a corresponding holographic
theory for a de Sitter brane embedded in AdS.
It would also be interesting to study the
detailed tensor structure \cite{csaba} 
of the propagator to see how robust the results
are. That there is no vDVZ discontinuity in de Sitter space 
\cite{Higuchi}-\cite{Kogan} is perhaps a good omen. 

\section*{Acknowledgments}
We thank Nima Arkani-Hamed, Andrey Barvinski,  Gia Dvali, Andrey Neronov, and
Ivo Sachs for helpful discussions. M. P. is 
supported by European Commission RTN Programme
HPRN-CT-2000-00131. S. S. is supported by the DFG Priority Programme SPP-1096.

\section*{Appendix}
In this appendix, we list a few useful formulas concerning associated 
Legendre functions.
The Legendre functions, $P^{\mu}_{\nu}(z)$ and $Q^{\mu}_{\nu}(z)$, that we
use are single-valued and regular for ${\rm Re} z > 1$, with a branch
cut from $-\infty$ to $+1$. They satisfy
\beq
(z^2 - 1) {d \over dz} P^{\mu}_{\nu} (z) = \nu z P^{\mu}_{\nu}(z) - 
(\nu + \mu) P^{\mu}_{\nu - 1} (z) \; , \label{derivative}
\eeq
\beq
(\nu - \mu + 1) P^{\mu}_{\nu + 1} (z) = (2 \nu + 1) z P^{\mu}_{\nu}(z) - 
(\nu + \mu) P^{\mu}_{\nu - 1} (z) \; , \label{identity}
\eeq
and identical equations for the $Q$'s.
Some closed-form expressions that are relevant are
\beq
P^{-3/2}_{3/2} (\xi) = {2 \over 3 \sqrt{2 \pi}} (\xi^2 - 1)^{3/4} \; ,
\eeq
\beq
P^{-3/2}_{1/2} (\xi) = {1 \over \sqrt{2 \pi}} \left[{\xi \over (\xi^2 - 1)^{1/4}}
- {1 \over 2 (\xi^2 - 1)^{3/4}} \ln \left ( {\xi + \sqrt{\xi^2-1} \over
\xi - \sqrt{\xi^2-1}} \right ) \right ] \; ,
\eeq
\beq
Q^{+3/2}_{1/2} (\xi) = -i {\sqrt{2 \pi} \over 4} {1 \over (\xi^2-1)^{3/4}} \; .
\eeq
We also need the asymptotics for large $\xi$. For $\nu > 0$, in the leading order we have
\beq
P^{\mu}_{\nu} (\xi) = {(2\xi)^{\nu} \over \sqrt{\pi}} 
{\Gamma\lf \nu + \half \rt \over \Gamma (\nu - \mu +1) } \lf 1 + {\cal O}
\lf {\ln \xi \over \xi^2} \rt \rt \; , \label{asymptP}
\eeq
\beq
Q^{\mu}_{\nu} (\xi) = {\sqrt{\pi} \over (2\xi)^{\nu + 1}} e^{i \mu \pi}
{\Gamma (\nu + \mu +1) \over \Gamma \lf \nu + {3 \over 2} \rt} \lf 1 + {\cal O}
\lf {\ln \xi \over \xi^2} \rt \rt \; . \label{asymptQ}
\eeq
The logarithmic subleading term appears in the ratio
\beq
{P^{iM}_{3/2}(\xi )\over P^{iM}_{1/2}(\xi )}
={2 \xi \over {3\over 2}-iM } 
\lf 1-{(1+4M^2)\over 8\xi^2}\ln \xi + {\cal O} \lf {1\over \xi^2} \rt \rt \; .
\label{ratio}
\eeq

\end{document}